\documentclass[12pt,preprint]{aastex}
\usepackage{epsfig}

\def\deg{\mbox{$^\circ$}}
\newcommand{\T}[1]{\mbox{$\times 10^{#1}$}}
\newcommand{\hms}[3]{\mbox{$\rm {#1}^h{#2}^m{#3}^s$}}
\newcommand{\dms}[3]{\mbox{${#1}\deg{#2}'{#3}''$}}

\def\kev{\mbox{ keV}}

\def\msol{\mbox{ $M_\odot$}}    % Solar Mass
\def\mdot{\mbox{$\dot{m}$}}	% accretion rate, rel to Eddington
\def\mdotcrit{\mbox{$\dot{m}_{\rm crit}$}}	% critical accretion rate
\def\Mdot{\mbox{$\dot{M}$}}	% accretion rate, absolute

\def\rtr{\mbox{$r_{\rm tr}$}}
\def\p{\mbox{p}}
\def\pmdot{\mbox{p$\dot m$}}

\def\ergscm{ergs~cm$^{-2}$~s$^{-1}$}

\def\ctss{cts~s$^{-1}$}

\newcommand{\simlt}  {\lower.5ex\hbox{\ltsima}}            % < over MMM

\newcommand{\simgt}  {\lower.5ex\hbox{\gtsima}}            % > over MMM

\catcode`\@=11 % This allows us to modify PLAIN macros.
\def\lsim{\mathrel{\mathpalette\@versim<}}
\def\gsim{\mathrel{\mathpalette\@versim>}}
\def\@versim#1#2{\vcenter{\offinterlineskip
        \ialign{$\m@th#1\hfil##\hfil$\crcr#2\crcr\sim\crcr } }}
\catcode`\@=12 % at signs are no longer letters

\def\etal{\mbox{\it et al. }}

\def\grs{GRS 1758--258}
\def\1e{\mbox{1E 1740.7--2942}}
\def\cygx1{\mbox{Cyg~X-1}}
\def\gratis{\mbox{\itshape GRATIS}}
\def\rxte{\mbox{{\itshape RXTE}}}
\def\rosat{\mbox{{\itshape ROSAT}}}
\def\cgro{\mbox{{\itshape CGRO}}}
\def\asca{\mbox{{\itshape ASCA}}}
\def\granat{\mbox{{\itshape GRANAT}}}
\def\sax{\mbox{{\itshape BeppoSAX}}}
\def\mir{\mbox{\itshape Mir}}

\begin{document}

\title{Long-term Multi-wavelength Observations of GRS~1758--258 and
the ADAF Model}
\shorttitle{\grs}
\shortauthors{Keck, \etal}
\slugcomment{for ApJ}

\author{
John W. Keck\altaffilmark{1}, 
William W. Craig\altaffilmark{2},
Charles J. Hailey\altaffilmark{1}, 
Fiona Harrison\altaffilmark{3},
Jae Sub Hong\altaffilmark{1},
Steven M. Kahn\altaffilmark{1},
Philip M. Lubin\altaffilmark{6},
Ryan McLean\altaffilmark{1},
Michael J. Pivovaroff\altaffilmark{1},
Michael Seiffert\altaffilmark{6},
Ron Wurtz\altaffilmark{2},
Klaus P. Ziock\altaffilmark{2}
}
\email{jwk@phys.columbia.edu}
\altaffiltext{1}{Columbia Astrophysics Laboratory, 538 W. 120th St., New York, NY 10027}
\altaffiltext{2}{Lawrence Livermore National Laboratory, 7000 East Avenue, Livermore CA, 94550}
\altaffiltext{3}{Space Radiation Laboratory, California Institute of
Technology, M.S. 220-47, Pasadena, CA 91125}
\altaffiltext{6}{Department of Physics, University of California,
Santa Barbara}

\begin{abstract}
We present a long-term multi-wavelength light curve of Galactic black
hole candidate \grs\ by combining previously published and archival
data from \granat, \rosat, \cgro, \rxte, \sax, \asca, {\it EXOSAT},
and the VLA.  In addition we include first spectral results from the
balloon-borne Gamma-ray Arcminute Telescope Imaging System (\gratis).
In light of divergent analyses of the 1991--1993 \rosat\ observations,
we have re-analyzed these data; we find the soft X-rays track the hard
X-rays, and that the fits require no black-body component---
indicating that \grs\ did not go to the high state in 1993.  We offer
an interpretation based on the ADAF model for a system with
$\mdot\lsim\mdotcrit$.  We find the 1990--1993 coeval hard and soft
X-ray observations support the ADAF predictions.  We discuss a new way
to constrain black-hole mass with spectral data and the ADAF theory,
and apply this technique to \grs\ to find $M_1 \gsim 8$--9\msol\ at an
assumed distance of 8.5 kpc.  Further investigations of the ADAF model
allow us to evaluate the model critically against the data and
flux-flux diagram of \citet*{BMG} and to understand the limits of the
latter's ``X-ray burster box.''
\end{abstract}


%\url{http://astro1.nevis.columbia.edu/gratis/}

\section{Introduction}
\grs\ is one of the most fascinating objects in the X-ray/gamma-ray
sky and we have much to learn about it yet.  For being second only to
\1e\ as the brightest persistent gamma-ray source near the Galactic
center \citep{1999ApJ...525..901M}, \grs\ remains an enigma.  Not only
do its remoteness and high column density obscure it, but its
proximity to GX~5--1, which delayed its discovery until 1990, even now
poses a minor obstacle to low-energy ($\lsim25\kev$) observations.

The few established facts about \grs\ are tantalizing.  The search for
a counterpart in the optical and infrared turned up no distinct
result, but only two or more candidates within $1''$ \citep{MARTI1}.
Even more interestingly, Rodr\'{\i}guez and collaborators working with
the VLA found not one, but at least three radio sources near the X-ray
source that have a roughly jet-like symmetry with respect to it
\citep{1992ApJ...401L..15R}.  This weak jet structure reveals \grs,
just as its neighbor \1e, to be a microquasar
\citep{1993A&AS...97..193M,1992IAUC.5477....1M}, but with a lower
column density than \1e \citep[cf.\ e.g.][]{1999ApJ...525..901M},
making it more amenable to observation.  \grs\ also shows similarities
to \cygx1, in that persistent, low-level hard emission dominates its
luminosity \citep{tanaka&lewin,1998PhR...302...67L}.  This persistence
puts \grs\ among the type III BHB's of \citet{zhang}, while the radio
jets put it among \citeauthor{zhang}'s type II BHB's.

In this paper we combine data from published and publicly available
data from high-energy missions and the VLA to produce a long-term
light curve for the source in the radio and in soft and hard X-rays.
We performed our own reduction of the VLA data, including several
observations without published results.  In the first part of the
paper (\S\ref{obs}), we enumerate the diverse data sources that we
have tapped, and our handling of them, especially of the 1990--1993
\rosat\ observations (\S\ref{rosat_data}), which are the only soft
X-ray observations bridging the 1991--1992 hard X-ray dip.

Second, we compare the data from the various data sets with each other,
and with results of observations of other celestial objects
(\S\ref{results}).  Most notably, we find (\S\ref{bb}) no ultra-soft
component throughout the twelve years spanned by the soft X-ray
observations, strongly arguing against the possibility that \grs\ was
in the high state at any of those times.  We investigate possible
correlations of hard X-rays with soft X-rays, and the hard X-rays with
spectral shape.

In \S\ref{dis} we combine the \rosat\ data with those of \granat\ to
constrain models of accretion disk emission: the ADAF model
(\S\ref{adaf.model}), the two-temperature model (\S\ref{tt.model}),
and the disk corona model (\S\ref{cc.model}).  Finally in
\S\ref{further} we show that the ADAF model explains properties of the
flux-flux diagram and ``X-ray Burster Box'' of \citet{BMG}, and
evaluate the model in light of the diagram and data of that paper.

\section{Observations\label{obs}}

In figure \ref{big_plot} we combine data from many instruments to
produce a long-term light curve and spectral index history for the
source. The soft X-ray light curve is of unabsorbed flux.

Below we briefly describe the major instruments and how the data were
derived.  Other instrument characteristics are listed in table
\ref{named}.  Minor, miscellaneous instruments are listed in table
\ref{misc}.

\subsection{\granat}

The ART-P and SIGMA telescopes on the Russian \granat\ satellite
discovered \grs\ in March--April 1990 \citep{1990IAUC.5032....1M}.
\granat\ observed the source seasonally through spring 1996.

The power-law indices in the plot come from table 1 of
\citet{1999AstL...25..351K}.  This table contains results of fitting
two seasonal spectra a year to a power-law over the range 30--300\kev.
Above 150\kev\ the spectrum decays exponentially.  In order to get
results comparable to other experiments' fits below 150\kev, we had to
band-limit the \granat\ photon indices.  We fit the spectrum from the sum
of their data (shown in their figure 5) below 150\kev\ in XSPEC to
$\alpha=1.825$, This index is 0.304 less than the power-law index for
the full 30--300\kev\ range.  We subtracted this number from each of
their table 1 indices to derive the band-limited index below
150\kev.  This procedure assumes the temporal independence of the
spectral shape, an assumption supported by the lack of evidence for
change in spectral hardness \citep{1999AstL...25..351K}.

The fluxes in the plot were produced by adding to the flux derived
from their 40--200\kev\ luminosities (their table 1), a power-law flux
from 20--40\kev, calculated using the just-derived band-adjusted
index.

\subsection{\rosat\label{rosat_data}}

\rosat\ observed \grs\ in September 1990, March 1992 and spring 1993,
spanning the hard X-ray dip of 1991--1992.  In section \ref{results},
we use these data to constrain the corona models for disk emission and
the ADAF model.  They are pivotal to our argument that there was no
ultrasoft component in 1993 and that the soft and hard X-rays
tracked through the 1991--1992 dip.

Two groups have published analyses of the \rosat/PSPC's 31 March--2
April 1993 pointed observation of \grs: \citet{mereghetti} and \citet[
Proc.\ 2nd Integ.\ Workshop.]{1997iwtu.conf..183G}. The original 1994
analysis found a high soft component and took source counts from a
radius of $18'$ around the source in an effort to collect
dust-scattered photons; this analysis took background counts from five
circles of radius $13'$ in the external regions of the detector to
avoid contamination by GX 5--1.

On the other hand, the 1997 re-analysis of the 1993 data showed a
relatively low soft component, and conjectured that the difference
comes from ``contamination of the field [of the original analysis] by
the nearby very bright source GX 5--1.'' For this analysis ``source
and background counts were extracted from a circle of radius $5'$ and
an annulus of radii $10'-15'$ (both centered at the \grs\ position)''
to exclude contamination by GX~5--1.

We conducted an independent analysis of the 1993 observation.  In
order to set the extraction radius for \grs\ we considered several
factors.  We wanted to include at least 90\% of the photons from the
source, taking into account the scattering halo \citep[assumed to
scale with that of GX 5--1, cf.\ ][]{1995A&A...293..889P} while firmly
excluding any photons from GX 5--1.  In addition we wanted to
eliminate any contribution from the brighter background sources, in
particular 1WGA J1800.2--2539 (localized only in 1995) at 13.5
arcminutes from \grs.  After examining several trial radii we
compromised on an extraction radius of nine arcminutes, with
background selected from a nine-arcminute radius circle at the western
edge of the detector.  To exclude any possible contamination from
GX~5--1, we extracted background from an area outside the central ring
of the PSPC \citep[cf.\ e.g.][]{2000AJ....119.2242C} rather than the
usual annular region just outside the central source region.

After extraction, the data were fit in XSPEC to a power-law with
photoelectric absorption and then to a black-body with absorption,
giving the respective best-fit values of $\alpha=1.75\pm0.39$,
$N_H=(1.28\pm0.13)\T{22}$ cm$^{-2}$ ($\chi^2_\nu=0.879, \nu=15$), and
$kT=0.53\pm0.06$, $N_H=(0.98\pm0.08)\T{22}$ cm$^{-2}$
($\chi^2_\nu=0.873, \nu=15$).  Figure \ref{rosat_pl} shows the
error contours of our fits.  It is noteworthy that the black-body fit
is inconsistent at more than the four-sigma level with the standard
column ($\sim1.5\T{22}\mbox{ cm}^{-2}$) for the source
\citep{1997ApJ...476..829M,1998ApJ...506L..35H,1999ApJ...525..901M}.
We will more fully address the blackbody fit in \S\ref{bb} below, but
suffice it to say for now that our fits indicate the black-body
component is incompatible with the accepted column for the source.

The count rate for the 7900-second spring 1993 observation was
$1.32\pm.02$ \ctss\ for 0.1--2.4\kev.  The 1990 September 10--12
%counts:      1.772    +/-1.50190E-02 cts/s (1-2.4 keV)
%   -bg:      1.155    +/-1.58674E-02 cts/s (1-2.4 keV)
%counts:      2.140    +/-1.65036E-02 cts/s (0.1-2.4 keV)
%   -bg:      1.317    +/-1.75296E-02 cts/s (0.1-2.4 keV)
All-Sky Survey observation of the source had a slightly higher count
rate of $1.81\pm.12$ \ctss\ for the same energy range.  This count
rate comes from the HEASARC \rosat\ All-Sky Survey Bright Source
Catalog, as revised in 1996, which supersedes the $0.95\pm0.04$ \ctss\
reported in \citet{mereghetti}.  The latter paper used preliminary
data to conclude that the ``soft and hard spectral components are
anti-correlated [if one ignores the 1992 observation].  Between 1990
September and 1993 March both varied by about a factor of 2, but in
opposite directions''.  Based on the more definitive count rate, we
find no anti-correlation, but rather a correlation between the 1990
and 1993 soft X-ray fluxes.  The HRI observation 13 March 1992
reported a rate of $0.15\pm0.01$ \ctss, statistically consistent with
the $0.18\pm0.01$ \ctss\ reported in \citet{mereghetti}.

\begin{figure}
\caption{Light curve and spectral index history for \grs\ for
1985-2000.  Provenance of miscellaneous data points is given in table
\ref{misc}.  Photoelectric absorption has been removed from the
low-energy X-ray fluxes by assuming the intrinsic spectrum is a
power-law that continues to lower energies.  Power-law indices with
errors greater than 0.4 have been omitted for clarity.  The dash-dot
lines indicate the ``burster-box'' boundaries of
\citet{BMG} for an assumed distance of 8.5 kpc.}
%\figurenum{1}
\label{big_plot}
\plotone{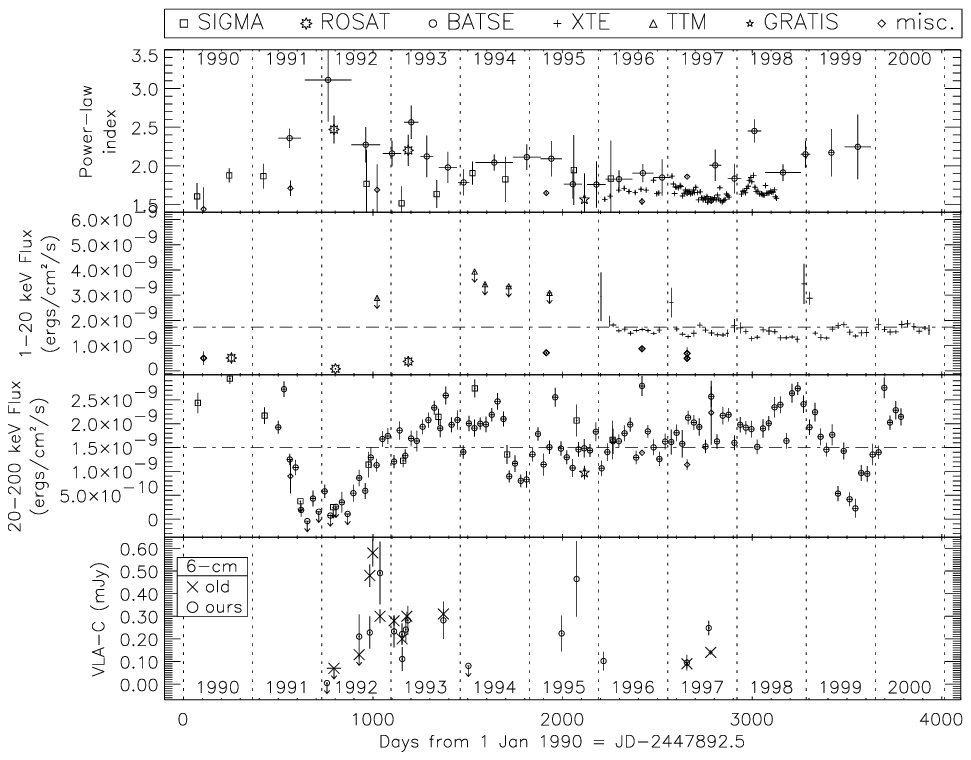}
\plotone{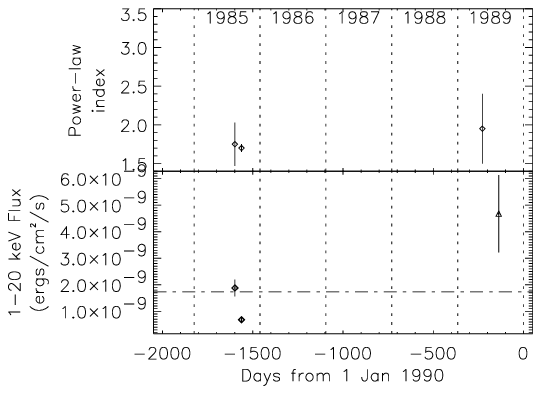}
\end{figure}

No spectral information is available from the 1990 \rosat\ All-Sky
Survey observation.  The HRI data provides no spectral information.
Thus we have assumed the 1993 spectral shape for all three
observations in deriving the unabsorbed fluxes in table
\ref{rosat_fluxes}, which is quite reasonable, given the large error
bars on our photon index, which easily encompasses the majority of
indices in figure \ref{big_plot}.

\citet{1997iwtu.conf..183G} found for the 4100-second 5 March 1992
off-axis PSPC observation of \grs\ a power-law index of $2.47\pm0.18$.
In contrast, we consider their flux for this observation to be
somewhat suspect since the proper extraction area is not unambiguous
in this distorted region of the focal plane.

In extrapolating \rosat\ fluxes for figure \ref{big_plot}, we assume
the photon index of 1993 held for 1990, but we use Grebenev's 2.47 for
the 1992 observation.

\subsection{\cgro}

\cgro/BATSE data comes from NASA's Marshall Space Flight Center: the
20--100 keV fluxes and their associated errors were produced by
fitting each day's BATSE Earth occultation data to a powerlaw with
spectral index 1.8.  To extend the energy range of the flux to
200\kev, we ignored the exponential cutoff above $\sim150\kev$ and
simply continued the power-law index of 1.8 out to 200 keV.  (We added
a 30\% correction to better agree with the \granat\ fluxes.)

Deriving the photon indices proved more of a challenge.  For each
viewing period we produced the \verb=.pha= and \verb=.rmf= files with
the standard FTOOLS \verb=bod2pha= and \verb=bod2rmf=.  We then fit
the data from each viewing period to a powerlaw within XSPEC and
recorded the results.  To give the values plotted in the figure, we
performed weighted averages of the spectral indices for each of three
periods within each observational cycle.  The errors on the indices
are large, but they hint at a trend resembling an upside-down cycloid,
with a cusp occurring during the hard X-ray dearth of 1991--1992, and
another in the dearth of 1999.

\subsection{\rxte}

The ASM fluxes presented in figure \ref{big_plot} are derived from
``quick-look'' results provided by the \rxte/ASM team and archived at
MIT.  These light-curves are binned into three energy bands,
1.5--3\kev, 3--5\kev, and 5--12\kev.  We converted these to fluxes by
comparing the count rates with the average ASM Crab count rates in
each channel, and extrapolated to 20\kev\ using a power-law index from
the last two bands.  The power-law indices ranged from $\sim1.7-2$ but
have been omitted from the PLI plot because of their large errors
($\sim0.8$).

Except for the two-month spike at the end of 1998-beginning of 1999,
the ASM data is flat not only in figure \ref{big_plot}, but also in the
individual bands.  There is no sign of the oscillation from late
February to September 1999 noted by
\citet{1999IAUC.7266....2S}.  It is difficult to tell if
the flatness is meaningful, or the result of inclusion of the Galactic
diffuse emission or other discrete sources in the ASM's large field of
view.  It might be possible to subtract the diffuse emission, but
unfortunately, the sky coordinates and position angle of each pointing
are not readily available.

\subsection{Other X-ray Data}

Data from the 10 April 1997 \sax\ observation were obtained from the
NASA's HEASARC archive and fit in XSPEC to a power-law with (MECS)
$N_H = (1.68\pm0.05)\T{22}$ cm$^{-2}$, $\alpha=1.65\pm0.02$,
$\chi^2_\nu = 1.17$ for $\nu=95$ and (LECS) $N_H =
(1.64\pm0.05)\T{22}$ cm$^{-2}$, $\alpha=1.54\pm0.02$, $\chi^2_\nu =
1.13$ for $\nu=107$.  For the PDS data, we found $\alpha=1.86\pm0.02$
with $\chi^2_\nu=2.19$ for $\nu=14$.

\gratis\ is a balloon-borne coded aperture imaging experiment
consisting of 36 co-aligned one-dimension telescopes with a total
effective area of 378 cm$^2$ at 30\kev.  Each telescope consists of a
two-mm-thick CsI(Na) crystal coupled to a photomultiplier tube (PMT)
passively collimated to have a field of view $40'\times 3\deg$ (FWZI).
The average energy resolution of the detectors is 19\% at 60\kev.
\gratis\ is more fully described in \citet{1989SPIE.1159...36H} and
\citet{instrument_paper}.

\gratis\ observed \grs\ from 5:43 to 6:43 hours UT on 17 October 1995.
\gratis\ was launched on its balloon from Alice Springs in the
center of the Australian outback at 23.80\deg~S, 133.40\deg~E.
%17.3991\mbox{ hours}, \delta=-37.8441\deg
To remove the background contribution, we subtracted the blank-field
pointing that immediately preceded this observation (4:43 to 5:43 UT)
centered at $\alpha = \hms{17}{23}{57}, \delta = \dms{-37}{50}{39}$,
epoch 2000.  This field contains no known high-energy sources.  \grs\
is at $l = 4.51\deg, b =-1.36\deg$ whereas the background pointing is
at $l=-9.93\deg, b=-1.03\deg$.  The two fields have comparable
contribution from the Galactic diffuse emission, since the
longitudinal variation is slow for $l<10\deg$
\citep{1982ApJ...260..111I}, and the latitudinal variation is nearly
flat when $|b|\ge 1\deg$ \citep[figure 2b]{1998ApJ...505..134V}.
GX~5--1 lay just on the edge of the fine-collimator field of view for
the source pointing.  In principle the rotations of some of the tubes
should permit photons from GX~5--1 to fall into the
coarse-collimators' $\sim3\deg$ field of view.  Omitting data counts
from these tubes did not significantly alter the 20--132\kev\ fit.

We unfolded the source counts using the standard XSPEC package from
NASA's HEASARC.  The response matrix was constructed with Lawrence
Livermore National Laboratory's COG Monte-Carlo photon transport
package \citep{COG2,COG1}.  A more detailed description of data the
analysis can be found in \S 4.1 of \citet{KECK}, available
online\footnote{http://www.astro.columbia.edu/\~{ }jwk/thesis.pdf}.
The best-fit power-law model to the data has a photon index of
$1.57\pm0.34$ with a reduced $\chi^2$ of 1.08 for nine degrees of
freedom.  The flux was ($9.3\pm1.4$)\T{-10} \ergscm.

The September 1985 {\it EXOSAT}/ME data came from the HEASARC archive
data for EXO~1757-259, with \grs\ 9.4 arcminutes off-center.  The
unfolded spectrum is plotted in \citet{SKINNER}.  The data file does
not list the region of the detector from which the source counts were
extracted, but the 45-arcminute field of view (FWHM) of the ME means
that the observation may include contamination from GX~5--1 in
addition to the Galactic diffuse emission.  Using XSPEC we fit the
data to a power-law with absorption: $N_H = (1.79\pm0.16)\T{22}$
cm$^{-2}$, $\alpha=1.70\pm0.05$, $\chi^2_\nu = 0.95$ for $\nu=61$.

The 1--20\kev\ flux point and upper limits for \mir-Kvant/TTM have been
extrapolated assuming a power-law index of 1.8.  Our data are taken
from the one detection and five upper limits of
\citet{1998AstL...24..742A}.

For completeness, we mention that the flux of $(0.862 \pm
3.36)\T{-10}$ \ergscm\ and power-law index of $1.95 \pm 0.45$ from the
15--180\kev\ POKER instrument \citep{1993A&AS...97..169B}.  The error
bar is far too large to add anything to the plot.

\subsection{Radio Data}

Radio fluxes measured by the Very Large Array (VLA) were originally
published in \citet{1992IAUC.5655....1M, MIRABEL93,
1992ApJ...401L..15R,2000ApJ...532..548L}\footnote{The 16 July 1992
flux quoted in \citet{MIRABEL93} supersedes that in
\citet{1992ApJ...401L..15R}, according to Luis Rodr\'{\i}guez and
Josep Mart\'{\i} (priv. comm., 2000).}.  The April 1997 data point was
provided by Josep Mart\'{\i}.

As part of our review of the data on \grs, we performed our own
reduction of the VLA observations of \grs, including those in 1994,
1995 and 1996, which have not been previously published.  Only in 1997
did the VLA begin observations of \grs\ at $\lambda=3.7$ cm, when
\citet{2000ApJ...532..548L} noted a flat spectral index.  We used the
classic, 15 October 1999 version of AIPS in our reduction.  For all
observations the gain calibrator was 1328+307 (a.k.a.\ 3C~286,
1331+305) and phase calibrator was 1748-253 (1751-253), except for the
3 August 1997 observation, which used 0134+329 (3C~48, 0137+331) for
the gain calibrator.  The phase calibrator had a boot-strapped flux
density in the range 0.47--0.52 Jy.  The upper limits in the plot of
these points are one-sigma, while those for the previously published
data are four-sigma.

The re-analyzed points are statistically compatible with the
previously published results, as can be seen in the bottom panel of
figure \ref{big_plot}.  The four additional points our analysis adds
to the light-curve add no further correlation to the X-ray data.

% 3 Aug 1997 (X Band) 1733-130 
\section{Results\label{results}}

We now further analyze and integrate the observations to compose a
picture of \grs.

\subsection{Black-body temperatures\label{bb}}

\citet{mereghetti} claimed to find a soft X-ray excess in 1993 data.
\citet{2000ApJ...532..548L} found that a fit to 1997 multi-instrument
observation with a power-law with exponential cutoff (PLE) was just as
good without the black-body ($\chi^2_\nu=1.0, \nu=620$) as with
($\chi^2_\nu=0.92, \nu=619$). \citet{1997ApJ...476..829M} found that
the fits to the \asca\ data did not require a black-body component as
the model of a lone power-law gave an acceptable ($\chi^2_\nu=1.031,
\nu=872$) fit.

Unlike the 1994 paper, but in agreement with the latter two papers, we
find that a black-body component is not needed in the \rosat\ data
and in the observations of other soft X-ray missions.

\begin{figure}
%\figurenum{3}
\caption{Confidence contours for fits to March 1993 fit to
\rosat/PSPC: {\it left} power-law fit\label{rosat_pl}; {\it right} black-body
fit\label{rosat_bb}.}
%\plottwo{rosat_pl.ps}{rosat_bb.ps}
\plottwo{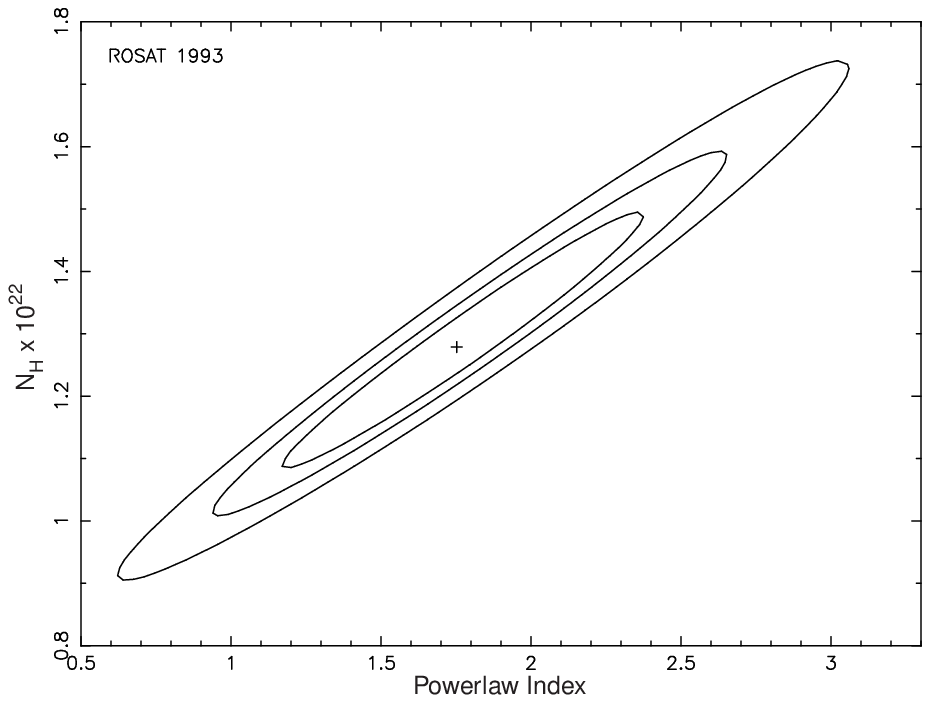}{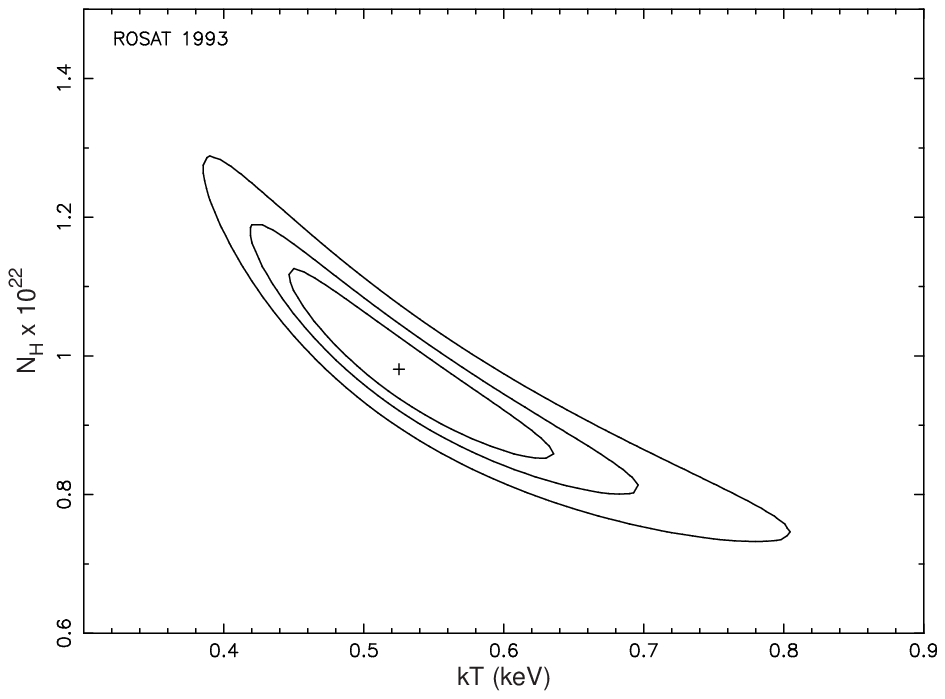}
\end{figure}

As we mentioned in \S\ref{rosat_data}, the typically observed column
density for \grs\ of $\sim1.5\T{22}$ cm$^{-2}$
\citep{1997ApJ...476..829M, 1998ApJ...506L..35H, 2000ApJ...532..548L}
lies far outside the three-sigma contour for our black-body fit to the
1993 \rosat/PSPC data, shown in figure \ref{rosat_bb}.  Our best-fit
column with this model ($N_H=(0.98\pm0.08)\T{22}$ cm$^{-2}$) is
inconsistent with the usual column at the four- or five-sigma level.
(While this could be an anomaly resulting from a sudden absence of
intrinsic absorption, we find this explanation unlikely in explaining
one highly deviant result.)  Additionally the power-law model provides
a good fit consistent with the standard column ($\chi^2_\nu \approx
0.97$, see figure \ref{rosat_pl}).

We also fit the data of other soft X-ray missions over twelve years.
The spectra from all of these observations are very similar to the
ASCA spectrum in figure 3 of \citet{1997ApJ...476..829M}.  Table
\ref{bb_temps} shows $\chi_\nu^2$'s that result with and without a
black-body component added to the powerlaw fit.  The respective
results are comparable, indicating that the ultrasoft component is
unnecessary.  As discussed above, the black-body component is
characteristic of the high state, so the lack of it indicates
that none of these observations occurred in that state.  Furthermore,
the absence of any major hard X-ray flux changes strongly argues
against the source changing states in this period.  Thus we conclude
that the source was not in the high state in this period.

\citet{1997ApJ...476..829M} estimated that to hide an ultrasoft
component behind a hot corona requires a relatively large scattering
opacity $\tau\sim2-5$.

As expected, the black-body temperatures fall in the range
$0.2\sim1.0\kev$.  For comparison, the black-body plus power-law fit
to the 1995 \asca\ data in \citet{1997ApJ...476..829M} found a
black-body temperature around 0.4--0.5\kev.  The best-fit HT model of
\citet{2000ApJ...532..548L} to the 1997 August {\it XTE} data gave a
black-body temperature $T_{\rm BB}\approx 1.18\kev$.

\subsection{High- and Low-energy Fluxes \& Flux-Flux Diagram\label{hilo}}

Whereas an anti-correlation between the hard and soft X-ray fluxes is
the usual sign of the transition between a low (hard) state and a
high (soft) state, the correlation of the \rosat\ data with the
\granat\ data through the 1991--1992 dip indicates that such a
transition did not take place.

\begin{figure}
%\figurenum{4}
\caption{Barret, McClintock, and Grindlay-style flux-flux diagram for
\grs\ at an assumed distance of 8.5 kpc.  The open plotting symbols
represent points whose soft and hard coordinates are only
approximately coeval. The ADAF model is scaled to $M_1=10.6\msol$ and
is discussed in section \ref{adaf.model}.  The dash-dot lines indicate
the ``burster-box'' boundaries of \citet{BMG}.
\label{bmg_grs}} 
\plotone{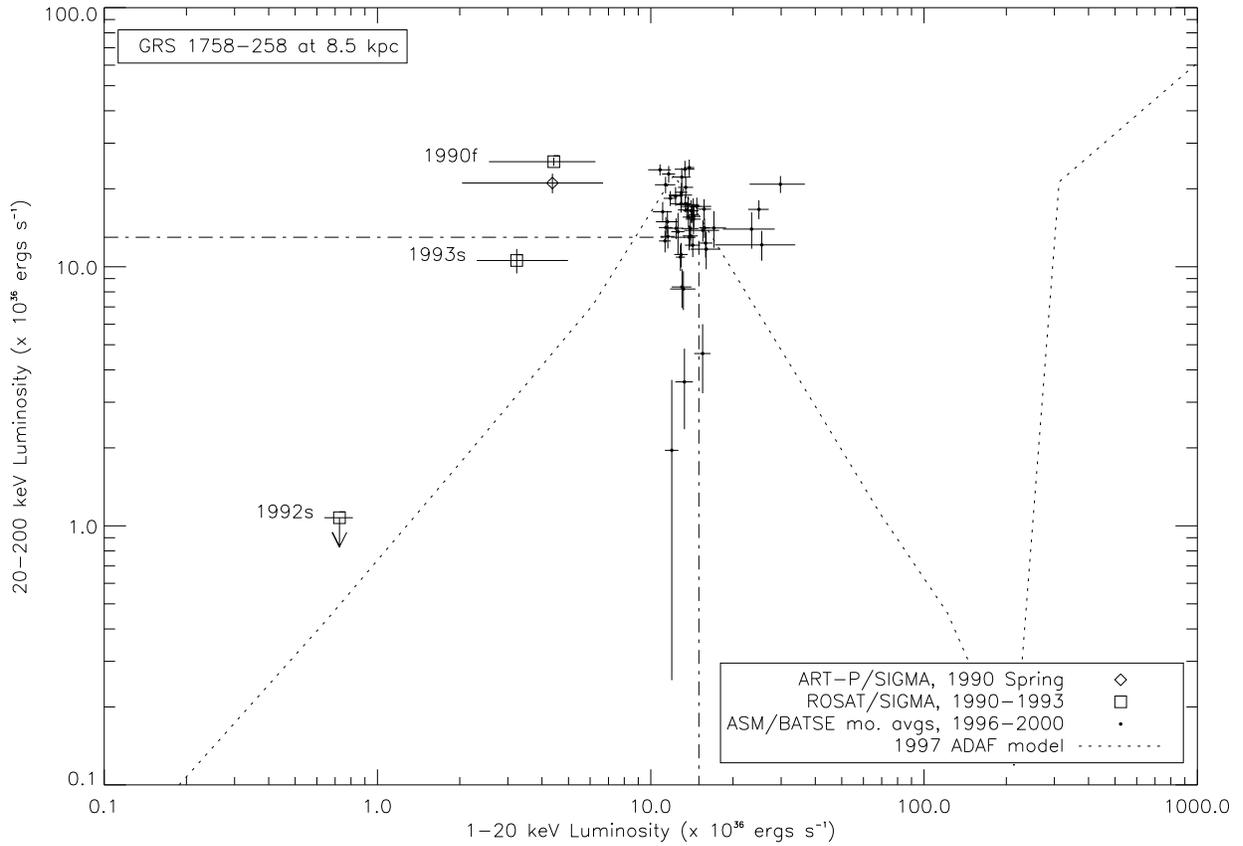}
\end{figure}

Figure \ref{bmg_grs} illustrates the correlations, or lack of, between
the hard X-ray (20--200\kev) luminosity and the soft X-ray (1--20\kev)
in the manner of \citet[hereafter BMG]{BMG}.

The \rxte/ASM data does not show any significant correlation, positive
or negative, to the hard X-ray dip in 1999.  Except for the two-month
spike at the end of 1998-beginning of 1999, the ASM data is flat
as shown in figure \ref{big_plot}, and in the individual bands.
There is no sign of the oscillation from late February to September
1999 noted by \citet{1999IAUC.7266....2S}.  It is difficult
to tell if the flatness is meaningful, or the result of inclusion of
the Galactic diffuse emission or other discrete sources in the ASM's
large field of view.  It might be possible to subtract the diffuse
emission, but unfortunately, the sky coordinates and position angle of
each pointing are not readily available.

\subsection{Correlation between Spectral Intensity and Shape
\label{shape-intense}}

Many black-hole candidates (e.g.\ \cygx1) display spectral pivoting,
so that the flux in a narrow energy band ($\sim10\kev$) remains
relatively constant in time, while the fluxes at higher and lower
energies anti-correlate.  This phenomenon occurs between the high and
low states, in what is known as the intermediate state.  A consequence
of spectral pivoting is that while the bolometric luminosity remains
relatively constant \citep{1997ApJ...477L..95Z}, a correlation is
established between the flux and power-law index in any given band
above or below the pivot.

The \granat\ photon indices are consistent with a constant spectral
slope because of the large error bars.  We see a suggestion of a
correlation between BATSE fluxes and spectral indices in figure
\ref{big_plot}.  For example, the flux dip of 1991--1992 has a higher
index and therefore a softer spectrum.

We wanted to test for a correlation between the photon index and the
flux, but calculation of flux assumes an index.  Such a method can
bias the results or exaggerate any correlation between intensity and
shape.

Hardness ratio and count rate are good metrics for photon index and
flux, and have the advantage of being independent of one another.  It
is difficult to compare raw count rates across instruments, so we
restrict our analysis to BATSE data, which has the longest temporal
baseline.

\begin{figure}
\caption{Hardness ratio vs. total count rate for 1990-2000 BATSE data.
The crosses are averages binned with boundaries at (A) 500 \& 1000,
(B) 1000 \& 1700, (C) 1700 \& 2600 (D) 2600 \& 3400, (E) 3400 \& 3700
days from 1 Jan 1990.\label{hard_corr}} 
\plotone{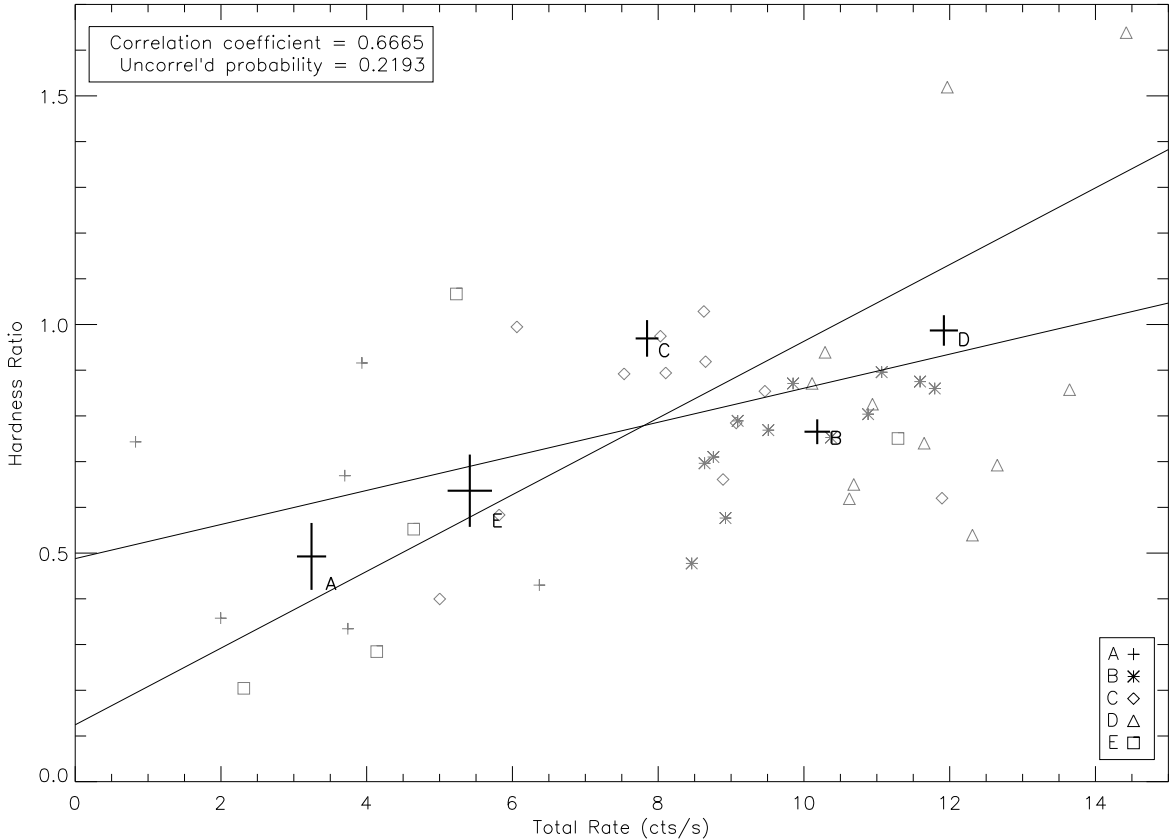}
\end{figure}

We plot the hardness ratio (ratio of counts above and below 78\kev)
against the total BATSE count rate in figure \ref{hard_corr}.  The
boundary between high and low counts is chosen to give a hardness
ratio $\sim1$ on average.  The thick crosses represent averages over
five periods that have similar total count rate throughout.  From the
slopes of linear fits to the averaged data, we find a
linear-correlation coefficient $r=0.67$ \citep[cf.][ pp.\ 198--200,
\S11.2]{BEVINGTON}.  The probability of an uncorrelated parent
distribution having a coefficient that big or bigger is only
$\sim22$\%, so there is a weak correlation between shape and
intensity.  Consequently we expect lower fluxes to have higher, softer
photon indices, a conclusion that agrees with the predictions of the
ADAF theory, discussed below (\S\ref{adaf.model}).

\section{Discussion\label{dis}}

Here we describe three models to account for the observed emission
from \grs.  In all of these models, the soft X-ray emission comes from
the disk and the hard X-ray emission comes from the corona or
hot-electron cloud.  The first is the advection dominated accretion
flow model, which we apply in a novel way.  The latter two are
standard Keplerian accretion models.

In our discussions of the latter two models, we assume the argument of
\citet*{SLE} that the seed photons range in energy from
0.05\kev\ to 5\kev.  (The exact mechanism of emission is not
important, but possibilities include black-body emission from the disk
and synchrotron radiation from disk electrons, as explained in
\citet{SLE,zhang}.)  In these models, we explain the
correlation between the hard and soft X-rays as a consequence of the
Comptonization of soft X-rays into hard by the corona.  The accurate
determination of the \rosat\ fluxes is critical here, as we
extrapolate the soft X-ray flux from that data.

\subsection{Advection Dominated Accretion Flow\label{adaf.model}}

We begin with the advection-dominated accretion flow (ADAF) model.
This is the most mature model of black-hole accretion, in that its
proponents have published spectral predictions
\citep{EMN,1998ApJ...505..854E}.  Thus we decided to test the model in
detail for \grs, and to scout out the theory-testing terrain that
imminent observational improvements will open to detailed exploration.

The accretion flow has two zones: an inner advection dominated flow,
and an outer thin accretion disk. The inner ADAF accretes into the
black hole as a quasi-spherical two-temperature corona, while the thin
disk is Keplerian.  The soft X-ray emission comes from the disk and
the hard X-ray emission comes from the corona.

The present model resembles the two-temperature model of \citet*{SLE}
with the difference that the transition radius between the
quasi-spherical and Keplerian zones, \rtr, changes with accretion
rate.  The most notable characteristic of the model is that the state
of black hole accretion, as illustrated in figure 1 of
\citet[hereafter EMN]{EMN}, is controlled by one parameter, the mass
accretion rate $\mdot \equiv \Mdot/\Mdot_{\rm Edd}$, where $\Mdot_{\rm
Edd}\equiv L_{\rm Edd}/0.10c^2$.  The last definition assumes a ten
percent radiative conversion efficiency.  (This assumption is
restricted to this definition; the radiative efficiency, $\epsilon$,
of the model itself varies with \mdot.)

In other words
\[ 
\mdot = \left(\frac{L}{\epsilon(\mdot)c^2}\right)
\left(\frac{0.10c^2}{L_{\rm Edd}}\right)
\]
or
\[
L \propto \mdot\epsilon(\mdot)L_{\rm Edd}
\]

Another significant feature of the ADAF theory is the scale invariance
of the states: the theory is the same for black holes of any size,
from Galactic to super-massive \citep{1995ApJ...452..710N}, which
means the radiative efficiency for a given \mdot\ is independent of
black-hole (primary) mass, that is $\epsilon \not = \epsilon(M_1)$.
Since the Eddington luminosity scales with the primary's mass, the
luminosity of any given state also scales with mass, thus establishing
a correlation between black hole mass and luminosity at known \mdot:
\[
L=L(\mdot) \propto M_1.
\]
If we know \mdot, we can compare the model luminosity at given mass to
measured luminosity to find the mass of the observed black hole
($M_1$).

We now use the mass-luminosity correlation and the observed flux
extremes to constrain the mass of \grs.  Since the ADAF models scale
with primary mass, the scale factor that best allows the model to fit
the data fluxes is the same factor that must multiply the model's
assumed mass.  In theory, the model's mass scaled by the factor is
then the mass of the primary.  In reality, other model parameters such
as the binary's inclination angle $i$, viscosity $\alpha$, and the
fraction of the total pressure due to gas (as opposed to magnetic)
pressure $\beta$ also effect the luminosity, though they do not effect
all energies and all states uniformly, as does the primary mass $M_1$.
Additionally, since the model predicts {\em luminosities} while we
measure {\em fluxes}, our technique only gives us the primary mass at
an assumed distance $d_0$, so instead of $M_1$, we find
$M_1/(d/d_0)^2$.

We take the ADAF model from the $\nu L_\nu$ curves of EMN for $M = 6
\msol$, $i=60\deg$, $\alpha = 0.25$, $\beta=0.5$ for Nova Muscae, and
of \citet[figure 1b]{1998ApJ...505..854E} for $M=9\msol, i=40\deg,
\alpha=0.3, \beta=0.5$ for \cygx1.  The luminosities of the high-state
spectra for the 1997 model are not exactly correct because the
high-state curves in figure 10 of EMN do not extend out to 200\kev.
To overcome this limitation, we have replaced the $\mdot=0.40$ curve
with its suitably renormalized counterpart from Figure 1a of
\citet{1998ApJ...505..854E}; the two curves agree to a constant factor
for their common energies.  The remaining high-state hard luminosities
are integrated from linear extrapolations, which may make them
slightly high, though not by much, judging from the extrapolation of
the $\mdot=0.4$ curve.

The 1998 model is an incremental improvement to the 1997 model.
Observational evidence from \citet{1997ApJ...488L.113Z} that the
transition radius varies in the low state and not exclusively in the
intermediate state moved the model's authors to incorporate this
feature.  Other changes reflect that a different source is being
modeled.  Figure 4a of EMN makes it clear that $i=30\deg$ instead of
$i=60\deg$ increases the flux $\sim25\%$ in both soft and hard X-ray
bands, at least for the low state.  The change in $\alpha$ is more
complicated, as its value effects the value of \mdotcrit.  For
$\alpha=0.25$, $\mdotcrit\approx0.082 $, while for $\alpha=0.3$,
$\mdotcrit\approx0.11$ and the flux shifts to harder energies at the
critical value, according to EMN's figure 4b.  The 1998 theory models
\cygx1, but the authors do not make clear how its wind-driven nature
effects the model.

\begin{figure}
\caption{Results of the simple calculation of the multiplicative factor
for the 1997 model.  The 1990 and 1997 \granat\ fluxes are taken from
figure 1 of \citet{1999AstL...25..351K}, and the power-law indices
from their table 1.  The one-sigma 1992 \granat\ upper-limit is
calculated from data of table 2 of \citet{1993ApJ...418..844G}.
\label{adaf_naive}}
\plotone{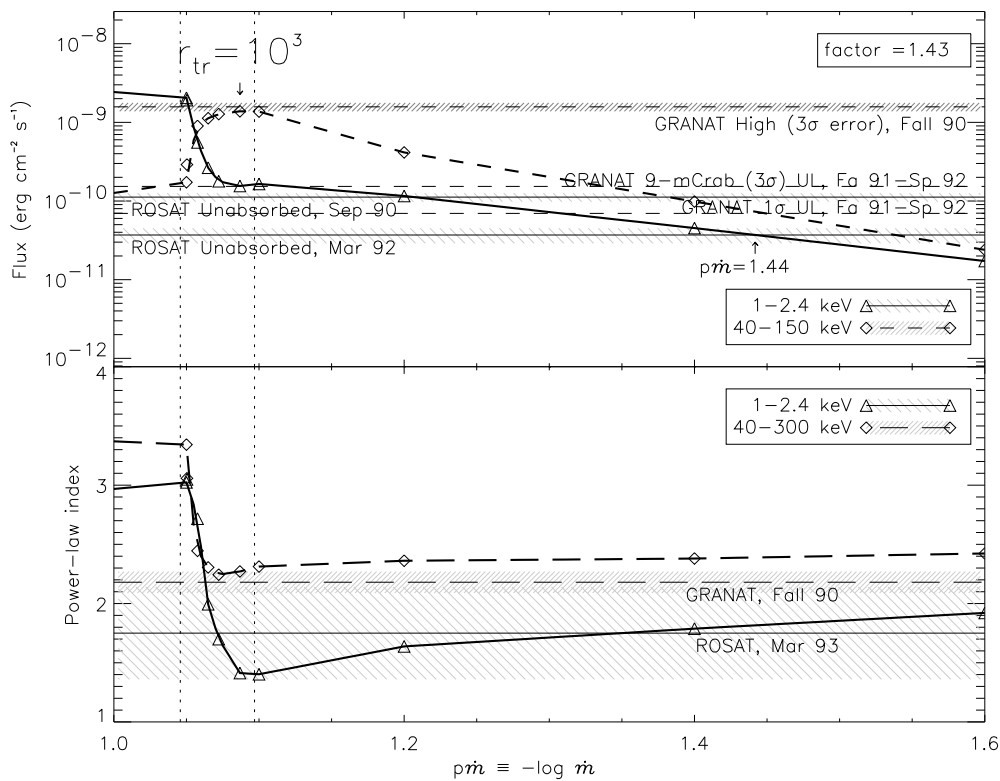}
\end{figure}

We find the minimum scaling factor for the model fluxes to produce the
maximum observed hard X-ray fluxes is 1.43.  For consistency, we can
check that the \granat\ and \rosat\ minima from the 1991--1992 dip are
still possible with this multiplicative factor; we can also check the
coeval photon indices.  Figure \ref{adaf_naive} shows the model
integrated over hard and soft energy bands as thick curves, along with
the observed flux extremes in those bands, plotted as thin horizontal
lines with hatched error bands.  The ordinate is
$\p\mdot\equiv-\log\mdot$, so higher \mdot\ is to the left.  We begin
in the low state on the right hand side.  Moving from right to left as
\mdot\ increases, we see both the hard ({\it dashed curve}) and soft
({\it solid curve}) fluxes increase monotonically until we cross the
first dotted vertical line, past which $\pmdot>\pmdot_{\rm crit}$,
putting the flow into the intermediate state.  The transition radius,
\rtr, rapidly contracts as we move to the left and the hard and soft
emission bands `swap' intensies.  As we further increase \mdot\ past
the second dotted line, the flow enters the high state, in which the
soft flux increases while the hard flux decreases.  The second panel
of figure \ref{adaf_naive} shows the power law indices predicted by
the model and observed by \granat\ and \rosat.  Note that the model's
hard X-ray power-law index increases with decreasing flux, in
agreement with the intensity--shape correlation we saw in the BATSE
data (\S\ref{shape-intense}).

We find that the theory is consistent with the \granat\ and \rosat\
observations.  The scaling factor multiplying the maximum hard model
flux is just enough to reproduce the maximum \granat\ flux in 1990 of
$92.5 \pm 3.8$ mCrab at the three-sigma lower-limit.  The factor also
allows the soft model flux to match the 1992 \rosat\ observed flux
minimum at the same \pmdot\ as the hard model flux falls below the
coeval \granat\ upper limit.  In the lower panel we see that the
model's hard photon index just brushes against the top error-bar of
the \granat\ 1990 index at approximately the same \pmdot\ at which the
1990 \granat\ max occurs.  At the \pmdot\ of the 1992 flux minima, the
model's soft photon index is well within the error bar of what we
assume to be the coeval \rosat\ index.  So we find that the ADAF
theory can consistently reproduce the \granat\ flux maximum and the
\rosat\ minimum at the same time as it predicts a hard X-ray flux that
falls below the \granat\ upper-limit.

Furthermore, if we attribute the multiplicative factor of 1.43 to
primary mass alone, then to agree with the \granat\ fall 1990 high at
the three-sigma level, we must have $M_1/(d/8.5\mbox{ kpc})^2 \gsim
8.6\msol$.\footnote{The 1998 paper samples spectra more sparsely in
\pmdot, so it is not clear what would be gained by a similar analysis
with it.}  Again we note that the spectral consequences of the assumed
values for inclination angle, viscosity and fractional gas pressure
are not clear from the published ADAF papers.

We could achieve a more sophisticated constraint on the factor by
running all of the \rosat\ and \granat\ flux and photon index data
through a chi-squared analysis, but the authors of the theory warn that
``the uncertainties in the model are still too large to draw
meaningful quantitative conclusions'' \citep{1998ApJ...505..854E}, so
such an analysis is unwarranted with the theory as it now stands.

This mass is similar to that found for \cygx1\, which is
interesting because \cygx1\ remains usually in the low state with only
occasional forays into the high.  Unlike \cygx1, which has a massive
companion and accretes via wind, the non-detection of an optical
companion for \grs\ constrains its mass to $\lsim4\msol$
\citep{1994ApJ...426..586C}, so it must accrete by Roche-lobe
overflow.  The other dynamically confirmed BHBs with identifiable
secondaries also accrete by Roche-lobe overflow, but most of them have
also been seen in a very high state (cf.\ figure \ref{renormd}, discussed in
\S\ref{further}).

\subsection{Comptonizing Donut Model\label{tt.model}}

%[include top diagram from figure 3 of SLE 1976]

While our primary goal was to study the ADAF model, we thought it
worthwhile to use our data with the two-temperature and corona models
to derive analytic estimates for some relevant parameters.

We apply the data to the two-temperature model of \citet*[hereafter
SLE]{SLE}.  The best fits to the spring 1993 observations data from
\rosat\ and \granat\ give $\alpha_{\rm soft}=1.8\pm0.4$, and
$\alpha_{\rm hard} \approx 1.5\pm0.2$.  We use these indices to
extrapolate the hard and soft fluxes measured by these missions to the
required energy bands: $F_{\rm soft,obs}(0.05-5\kev)=
(3.4\pm1.6)\T{-10}$ \ergscm\ and $F_{\rm hard}(5-150\kev) =
(1.30\pm0.19)\T{-9}$ \ergscm.  So $L_{\rm soft,obs}/L_{\rm hard} =
0.26\pm0.13$.

We use these fluxes to derive model parameters.  Combining the model's
assumed cloud temperature of $\sim50\kev$ with $\alpha_{\rm hard}$, we
obtain $\tau\approx2.0$.  Our result agrees with others.
\citet{1999AstL...25..351K} fit the 1990--1997 \granat\ data to a
Comptonized disk model and found $\tau\approx1.2$ and
$T\approx41\kev$.  on the other hand the best-fit HT
\citep{1994ApJ...434..570T,1995ApJ...449..188H} plus black-body model
of \citet{2000ApJ...532..548L} to the 1997 August {\it XTE} data gave
$\tau\approx3.4$ with an electron (cloud) temperature
$T\approx52\kev$.

We likewise find from our data the scattering fraction $\zeta \approx
(20\pm 12)\%$.  So less than a third of the soft X-rays need be
upscattered by a Comptonizing cloud that pushes the bounds of optical
thinness to give the hard X-ray ($\gsim5\kev$) spectrum in spring
1993.  This result is consistent with the $\sim10$\% SLE found by
applying their model to \cygx1.  Our simple analytic results agree
with the broad conclusions of others concerning general model
parameters.

\subsection{Partial Disk Corona\label{cc.model}}

We now consider the disk corona model first proposed by
\citet{1977ApJ...218..247L} and \citet{1977A&A....59..111B} as an
alternative to the two-temperature model that would stabilize the
disk.  In this model the corona is adjacent to the disk and sandwiches
it from above and below like an undersized hamburger bun.  See
\citet{1998PhR...302...67L} for further discussion.

As with the previous model, the disk emits soft X-rays
$L_{\rm soft}$ which are then Compton up-scattered to produce the hard
X-rays $L_{\rm hard}$.  The corona covers some fraction, $\xi$, of the
disk, so $(1-\xi)$ of the original photons have no chance of
interacting with it.  If the corona's optical depth is $\tau$, another
$\xi e^{-\tau}$ of the original photons pass through the corona
unaffected.  The formalism for deriving the covering fraction in terms of
observables is identical to that of the previous model, with
$\zeta=\xi(1-e^{-\tau})$.  The covering fraction is then
\[
\xi = \frac{1}{1-e^{-\tau}} \left(1+\frac{A L_{\rm soft,obs}}{L_{\rm hard}}\right)^{-1}.
\]

The extrapolated luminosity ratio ($150 : 5.0 : 0.05\kev$) from the
spring 1993 \rosat\ and \granat\ data remains $L_{\rm soft,obs}/L_{\rm
hard} = 0.26\pm0.13$.  From the \granat\ spring 1993 Comptonized disk
model temperature of $\approx 85\kev$ and $\alpha_{\rm hard} =
1.5\pm0.2$, we find $y\approx1.6$ and $\tau\approx2.4$. For this $y$,
we get the same amplification factor as for the previous model,
$A\approx15\pm5$, so the covering fraction $\xi \approx
\zeta/(1-\exp(-\tau)) =9$--35\%.  So less than about a third of the
emitting region of the disk was covered by a corona in spring 1993.

Our purely spectral analysis provides independent confirmation of
results from timing data.  \citet{2000ApJ...531..963L} found that the
lack of PDS steepening with increasing photon energy eliminated the
$\zeta\sim1$ model investigated by \citet{1998ApJ...506..281B}.
\citet{maccarone} found a $\sim25$\% reprossessing fraction by
interpreting the soft X-ray lags in XTE~J~1748--2848 with a radiative
feedback model.

\section{Further Inquiries with the ADAF Model\label{further}}

We now explore the ADAF model in the context of the flux-flux
diagram of \citet*[BMG]{BMG}.  We find that the ADAF model explains the
boundaries of the ``burster-box'' and reveals a gap in the diagram's
differentiation of black-holes from neutron stars.

\begin{figure}
\caption{Plot of \citet[cf.\ BMG]{2000ApJ...533..329B}.  The neutron
star luminosities are unabridged, but the black hole luminosities have
all been renormalized to 3\msol.  Overplotted for comparison is the
ADAF model of \citet{EMN} and \citet{1998ApJ...505..854E} normalized
to that same mass.  Small unfilled circles are based on the second
distance estimate in BMG.  The numbers in the polygons are the model
\mdot's at each point.  The dash-dot lines indicate the
``burster-box'' boundaries of \citet{BMG}.
\label{renormd}}
\plotone{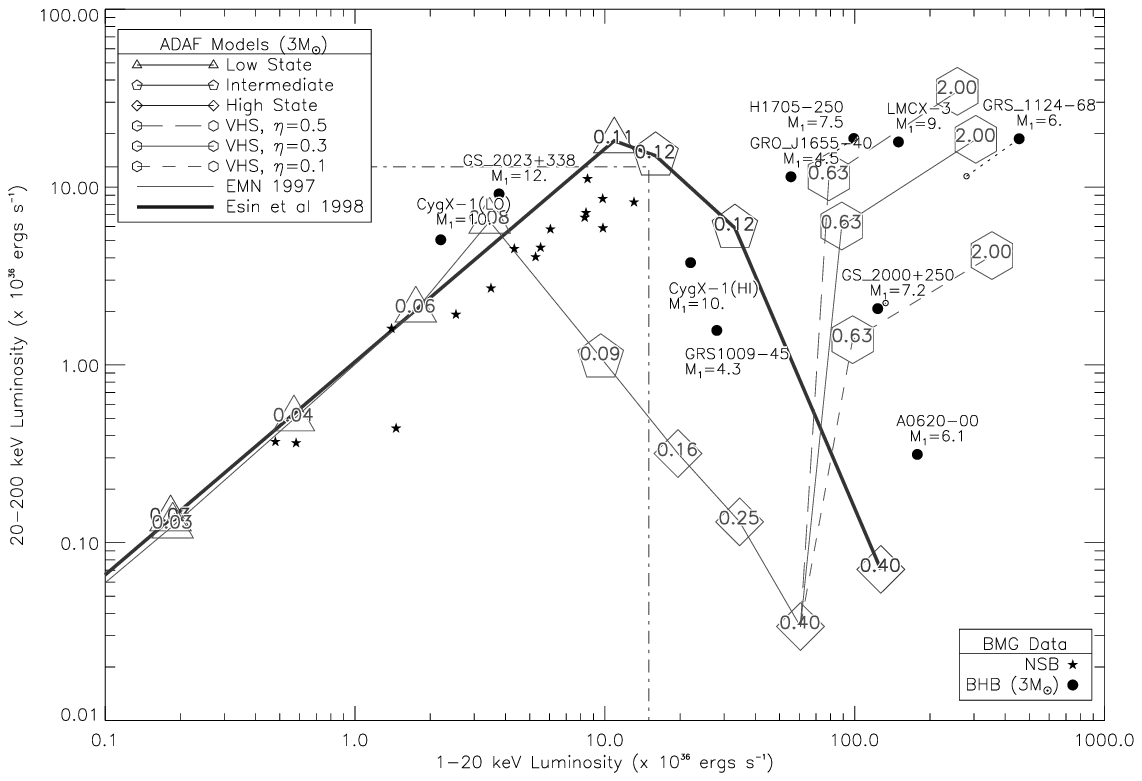}
\end{figure}

In figure \ref{renormd} we re-present figure 13 of
\citet{2000ApJ...533..329B} overlaid with the tracks of the ADAF
models of \citet[EMN]{EMN} and \citet{1998ApJ...505..854E}.  To make
the comparison meaningful, we have scaled the luminosities of each
black hole binary by $3\msol/M_1$, and the respective 1997 and 1998
models by $3\msol/6\msol$ and $3\msol/9\msol$. So assuming scale
invariance of the ADAF model holds, we normalize the luminosities to
$M_1=3\msol$.

With one exception, all of the black hole binaries (BHBs) are in
reasonable agreement with the theory (though, since BMG's data lacks
error bars, it is difficult to tell what reasonable means).  The solid
points show the BHBs at their highest observed hard luminosities, so
it is no surprise that the plot catches them either at the corner that
represents \mdotcrit\ or in the very-high state. The 1996 paper flags
A0620-00 as having a somewhat doubtful hard X-ray tail.  With that one
exception, the models appear to have good agreement with the theory,
including the very high state theory.  No one value for the fraction
of disk energy dissipated directly in the corona, $\eta$, clearly
stands out, but $\eta=0.5$ is reasonably close to three BHB points,
whereas the other two values of $\eta$ have only one each.  It is
remarkable that the 1998 theory, which was constructed for \cygx1,
predicts soft X-rays that are not only over-luminous in both low and
high states, but that are both over-luminous by the same factor of
$\sim1.8$.  A similar factor obtains for GRS~1009-45, whose mass has
only recently been measured \citep{1999PASP..111..969F}.

Remarkably the 1998 theory places the transition between low and
intermediate states, which is the first simultaneous maxima of soft
and hard X-rays, at the corner of the burster box.  Notice that the
neutron-star binaries (NSBs), which have not been renormalized by
mass, lie along the theory's low-state line (which runs roughly
parallel to the $y=x$ line along which the luminosity points scale
with $Md^2$).  Consequently the BMG plot does not distinguish {\em
low-state} BHBs from NSBs.  Since NSBs would likely show a black-body
component \citep{1997ApJ...478L..79N} and this component appears in
BHBs only with the onset of the intermediate and high states, perhaps
the presence or absence of the black-body component would distinguish
NSBs from BHBs within the box.  This criterion falls short if the 1997
model is the more accurate, because the model predicts that
intermediate-state and some high-state black holes fall in the box.

\begin{figure}
\caption{Data from \citet*[BMG]{BMG}, figure 3.  Overplotted for comparison
are the ADAF models of \citet*[EMN]{EMN} and \citet{1998ApJ...505..854E}
renormalized to $M_1=3\msol$. The state names are those of BMG and are
somewhat different than in the ADAF model. \label{gx339}}
\plotone{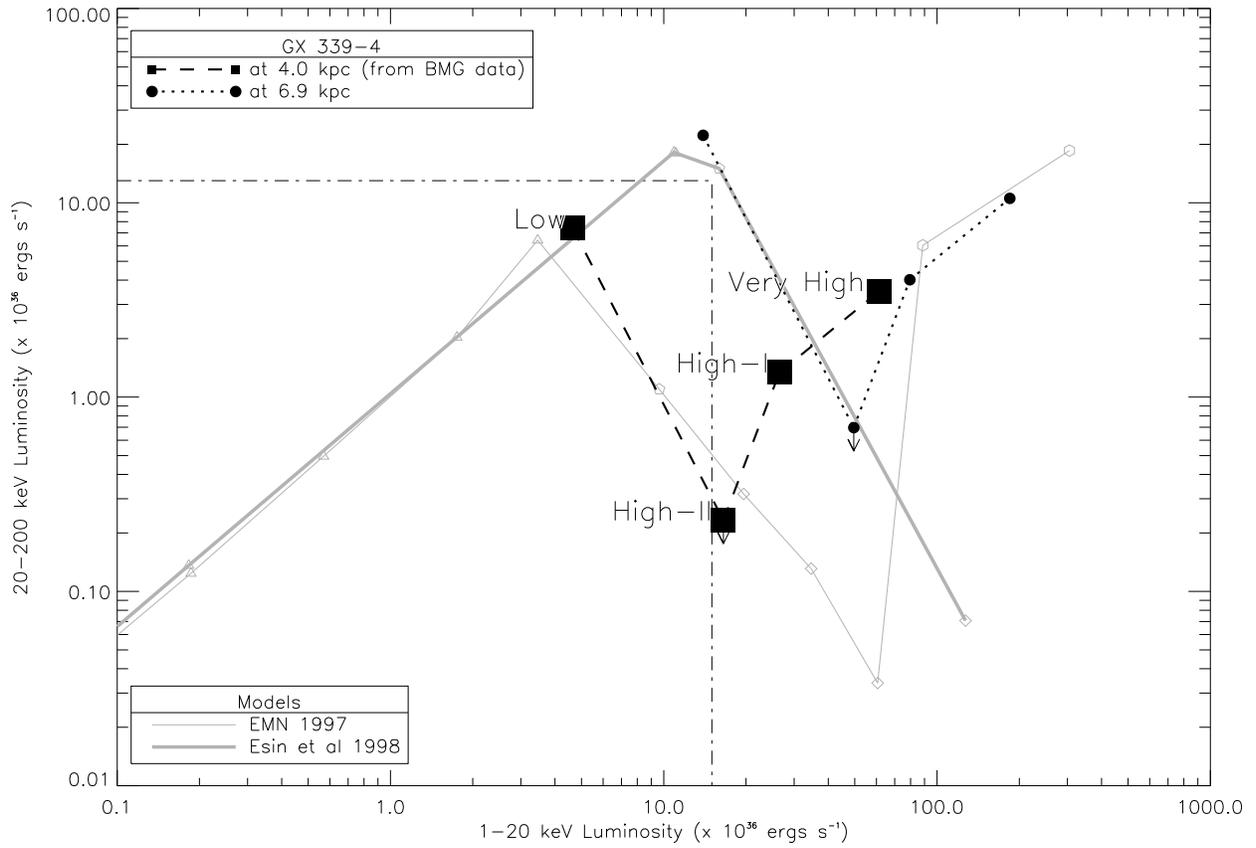}
\end{figure}

BMG's figure 3 shows the plot for four observations of GX~339-4 at an
assumed distance of 4.0 kpc.  The points in the plot are labeled from
left to right ``Low state,'' ``High state II,'' ``High state I,''
``Very high state.''  (The nomenclature is apparently different from
that of the ADAF papers.) In figure \ref{gx339} we compare this data
to the ADAF models at minimal ($3\msol$) black-hole mass.  We see that
the high-state segment of the 1997 $\alpha=0.25$ model agrees
reasonably well with the data, but that the ``high-I data point''
(third from left) is somewhat distant from both the high and the
very-high state segments.  EMN caution that their very-high state is
speculative.  \citet{1998ApJ...505..854E} do not treat this state for
Cyg~X-1, so we do not know how the very-high-state theory changes with
the change in parameters between the papers, but it is safe to assume
that it cannot change without losing agreement with the BHB points on
the BMG plot of figure \ref{renormd}.  This part of the theory comes
reasonably close to the data, as well as giving good agreement between
the data and the 1998 $\alpha=0.30$ model for the low and high states,
on the assumption that GX~339--4 is a factor of $\sim\sqrt{3}$ more
distant, in closer proximity to the Galactic center ($\sim7$ kpc).
The typical absorption column to the source \citep[1--$9\T{21}$
cm$^{-2}$;][]{1999ApJ...519L.159B,1997ApJ...479..926M} is consistent
with this conclusion.  Of course, any agreement with the ADAF theory
presumes that the source is in fact a black hole, a premise which is
not without dissent \citep[e.g.][]{1987AJ.....93..195C}.  The same
authors estimate the 4.0 kpc distance to the source from its color and
by assuming its interstellar line velocity is due to differential
rotational in the Galactic plane.

\begin{figure}
\caption{Data from \citet*{BMG}, figure 2.  Underlaid for comparison
are the ADAF models of \citet*{EMN} and \citet{1998ApJ...505..854E}
renormalized to the mass of Nove Muscae, $6\msol$.  The numbers
enclosed in the circles represent the day of the year 1991 for each
point; the dotted line is only to guide the eye.\label{nova_muscae}}
\plotone{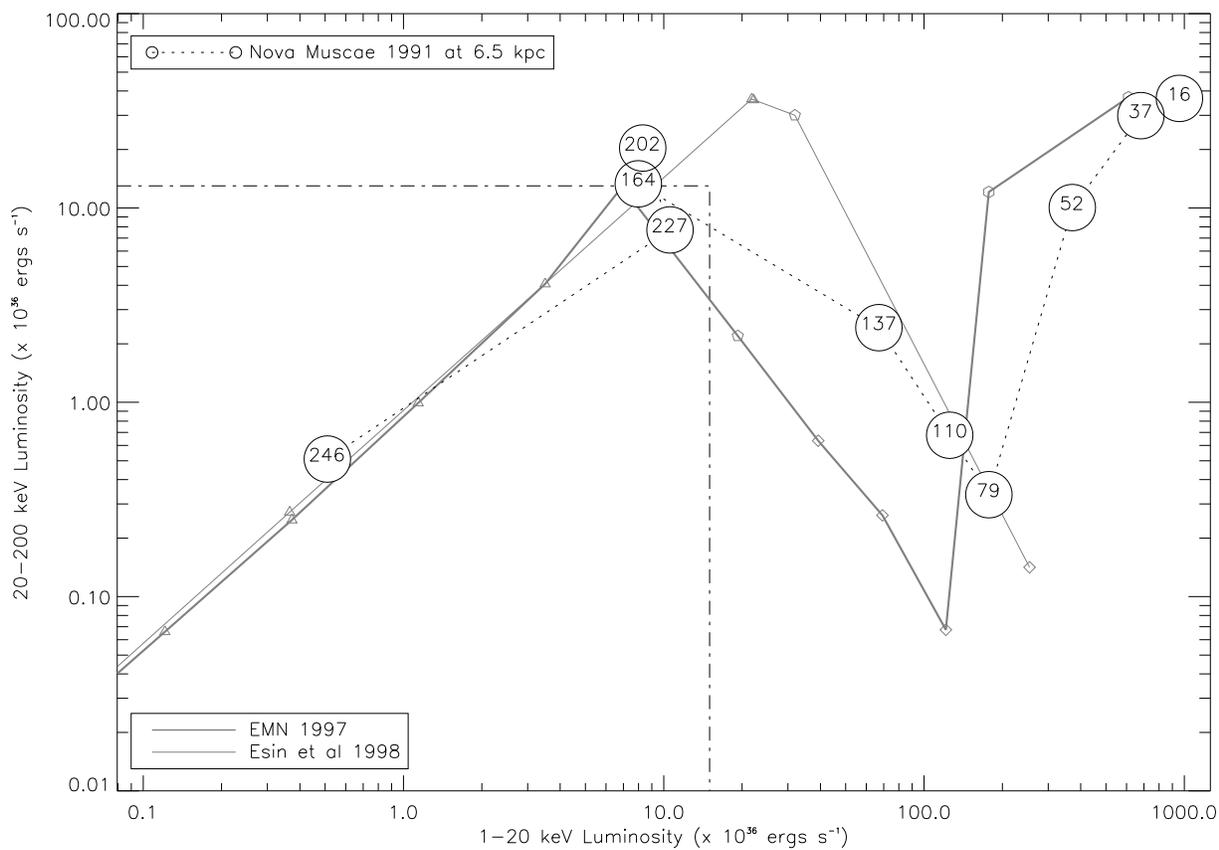}
\end{figure}

BMG's figure 2 shows the plot for ten points from the burst of Nova
Muscae 1991 (GRS~1124-68) after its X-ray maximum 16 January of that
year.  In figure \ref{nova_muscae} we compare this data to the ADAF
models at ($6\msol$) black-hole mass.  Recall that the model of EMN
(1997) was constructed to fit this source.  It is not surprising that
the EMN model fails to fit the very high state data, since the model's
authors ventured only a tentative proposal for that state.  What is
surprising is that the high-state model is so far removed from the
data, as figure 12 of EMN primed us to expect that the model was in
much better agreement with this state.

\section{Conclusion}

We have seen that the preponderance of the data indicates that \grs\
was in a low state in the early 1990's, since there was no black-body
component (\S\ref{bb}) and the soft and hard X-rays were correlated
(\S\ref{hilo}).  (Lack of dependable soft X-ray data prevents us from
speaking about the 1999 X-ray dearth.)  We found a correlation between
the hard X-ray spectral intensity and shape.

We applied the data to three models of accretion disk emission and
used it to constrain the geometries of the Comptonizing torus
(\S\ref{tt.model}) and the disk-corona models (\S\ref{cc.model}).  In
the former, two-temperature model, less than a third of the soft
photons are upscattered to give the observed hard emission, while the
latter disk-corona model would say that about the same fraction of the 
disk is covered by a $\tau\approx 2$ corona, consistent with the
$\sim10$\% reprocessing fraction found by SLE in applying their theory
to \cygx1.

In \S\ref{adaf.model} we found that the coeval \granat\ and \rosat\
data were consistent with the ADAF theory for an accreting black hole
that spends most of its time in the low state, but perhaps
occasionally sidles up into the intermediate state.  The correlation
between the hard X-ray intensity and spectral shape observed in the
BATSE data (\S\ref{shape-intense}) further supports the theory.  We
then introduced a new technique for constraining the mass of the
primary and applied it to \grs.  With the present data we constrained
the mass of the black hole primary $M_1/(d/8.5\mbox{ kpc})^2 \gsim
8.6\msol$.

In \S\ref{further} we critically assessed the ADAF model against the
flux-flux diagram of \citet[BMG]{BMG}.  We found that the ADAF model
explains the high-luminosity corner of the ``burster-box'' but also
raises the possibility that the box includes not only the advertised
neutron stars, but also some low-luminosity black holes.  Furthermore
we found differences between the ADAF model and BMG's data for Nova
Muscae 1991 and GX~339-4.  The discrepancy for the latter appears to
be overcome by positing a distance of $\sim7$ kpc instead of 4.0 kpc.

For a more critical test of the ADAF theory along this same line,
additional data from binaries in various states is necessary.
Continuous all-sky monitoring in both the soft and hard X-ray bands,
such as would be provided by EXIST
\citep{exist99,2000HEAD...32.2004G}, would be key to this effort, and
the luminosity's mass-scaling allows us to compare directly the tracks
of black holes of diverse masses across a flux-flux diagram such as
BMG's.  (As many BHC's have exponential tails $\gsim100\kev$
[\citeauthor{1998PhR...302...67L} \citeyear{1998PhR...302...67L}],
lowering the upper limit of the hard flux of the diagram to
$\sim100\kev$ could well improve its convenience and resolving power
in comparing BHC's.)  Timing signatures
\citep[cf.][]{1999ApJ...519L.159B} could then establish the
corresponding accretion state.

\acknowledgements{This research was partially funded by NASA's
Graduate Research Fellowship, NASA SRT, Lawrence Livermore (LLNL) IRD
and DOE NN-20.  Work performed at the University of California
Lawrence Livermore National Laboratory is supported by the US
Department of Energy under contract W7405-ENG-48.  It has made use of
data obtained through the High Energy Astrophysics Science Archive
Research Center Online Service, provided by the NASA/Goddard Space
Flight Center, and extensively used NASA's Astrophysics Data System
Abstract Service.  We thank Michael McCollough and Colleen
Wilson-Hodge of MSFC for graciously providing the \cgro/BATSE data,
and Josep Mart\'{\i} for the the April 1997 VLA data point.
The unseen but indispensable hands behind \gratis\ include Dennis
Carr, Chris Adams, Todd Decker, Jim Hughes, Greg Sprehn, Leigh
Brookshaw, Craig Brooksby, Bob Priest, Matt Fischer, and Irwin
Rochwarger.  Thanks to David Helfand for his help negotiating the
tortuous AIPS package, and to the National Scientific Ballooning
Facility for making the \gratis\ flight possible.  We thank the
referee for the promptness of his comments.}

\bibliography{bib}
\bibliographystyle{apj}

\begin{deluxetable}{lcl}
\tablecaption{Instruments named in figure 1\label{named}}
\tablehead{
\colhead{instrument} & \colhead{band (keV)} & \colhead{citation}
}
\startdata
\granat/SIGMA & 30--1300    & see text, cf.\ \citet{1999AstL...25..351K} \\
  \rosat/PSPC & 0.1--2.5    & this work\\
  \rosat/HRI  & 0.1--2.5    & this work \\
 \cgro/BATSE  & 20--1000    & this work\\
   \rxte/PCA (PLI)  & 2--60       & \citet{1999ApJ...525..901M}, fig. 2\\
   \rxte/ASM (flux) & 1.5--12     & this work\\
\mir-Kvant/TTM &  1--20      & see text, cf.\ \citet{1998AstL...24..742A} \\
     \gratis  & 20--200     & this work\\
\enddata
\end{deluxetable}

\begin{deluxetable}{rrlcl}
\tablecaption{Miscellaneous data points in figure \ref{big_plot}.\label{misc}}
\tablehead{
\colhead{mo. year} & \colhead{day\tablenotemark{a}} 
& \colhead{instrument} 
& \colhead{band (keV)} 
& \colhead{citation}
}
\startdata
Aug 1985 &  -1599 &    SL2/XRT     & 2.5--25  & \citet{SKINNER} \\
Sep 1985 &  -1562 &{\it EXOSAT}/ME & 1.5--50  & this work \\
May 1989 &   -228 &      POKER     & 15--180  & \protect{\citet{1993A&AS...97..169B}} \\
Apr 1990 &    106 &\granat/ART-P   &  4--60   & \citet{SUNYAEV1990}\\
Jul 1991 &    564 & \cgro/OSSE     &50--$10^4$& \citet{JUNG93} \\
Oct 1992 &   1021 &\mir-Kvant/HEXE & 20--200  & \citet{MAISACK} \\
Mar 1995 &   1914 &   \asca/SIS    & 0.4--10  & \citet{1997ApJ...476..829M}\\
Aug 1996 &   2419 &\rxte/PCA-HEXTE & 2.5--250 & \citet{1998ApJ...506L..35H} \\
Apr 1997 &   2657 &   \sax/LECS    & 0.1--10  & this work \\
Apr 1997 &   2657 &   \sax/MECS    & 1.3--10  & this work \\
Apr 1997 &   2657 &    \sax/PDS    & 15--300  & this work \\
Aug 1997 &   2784 & HEXTE\&OSSE    & 20--700  & \citet{2000ApJ...532..548L}, HEASARC archive \\
\enddata
\tablenotetext{a}{referenced to 1 Jan 1990 = JD 2447892.5}
\end{deluxetable}

\begin{deluxetable}{lccl}
\tablecaption{Unabsorbed 1.0--2.4\kev\ Fluxes from \rosat \label{rosat_fluxes}}
\tablehead{
\colhead{} & \multicolumn{2}{c}{$10^{-10}$ ergs cm$^{-2}$ s$^{-1}$} \\
\cline{2-3} \\
\colhead{Observation Date} & 
\colhead{Original\tablenotemark{a}} &
%\colhead{Re-analysis\tablenotemark{b}} &
\colhead{Our Re-analysis} & 
\colhead{Instrument}
}
\startdata
1990 Sep 10--12 & 1.4 		& $1.12\pm0.14$ & PSPC (survey)\\
1992 Mar 13 	& 0.76 		& $0.51\pm0.06$ & HRI\\
1993 Mar 31--Apr 2 & 2.8 	& $0.82^{+0.11}_{-0.08}$ & PSPC \\
\enddata
\tablenotetext{a}{from \citet{mereghetti}}
%\tablenotetext{b}{calculated from fit parameters and absorbed flux 
%of \citet{1997iwtu.conf..183G}}
\end{deluxetable}

\begin{deluxetable}{rlllrrll}
\tabletypesize{\scriptsize}
\tablecaption{Black-body temperatures\label{bb_temps}}
\tablehead{
\colhead{date} & 
\colhead{BB temp.} & \colhead{BB norm} & 
\colhead{model\tablenotemark{a}} & \colhead{$\chi_\nu^2$} & 
\colhead{$\chi_\nu^2$ w/o BB} & 
\colhead{instrument} & \colhead{notes}
}
\startdata
Sep 1985 & $0.2370 \pm 0.6229$ & $6.0\T{-4}$ & BB+PL & 
	1.095 (32) & 1.037 (34) & {\it EXOSAT}/ME & $\alpha=1.70\pm0.10$\\
Sep 1985 & $1.063 \pm 0.1112$ & $9.8\T{-4}$ & BB+PL & 
	1.048 (32) & 1.037 (34) & {\it EXOSAT}/ME & $\alpha=1.32\pm0.41$\\
Sep 1985 & $0.2603 \pm 0.09307$ & $1.7\T{-3} $& BB+CompST &
	0.993 (57) & 0.985 (38) & {\it EXOSAT}/ME &\\
Mar 1993 & $0.5734 \pm  0.4028$ & $1.0\T{-3}$ & BB+PL &
	0.998 (13) & 0.879 (15) & \rosat/PSPC &\\
Mar 1993 & $0.6488 \pm  1.034$ & $7.9\T{-4}$ & BB+CompST &
	1.091 (12) & 0.947 (14) & \rosat/PSPC &\\
Mar 1995 & $0.5374 \pm 0.0265$ & $7.1\T{-4} $ & BB+PL & 
	0.971 (280) & 1.051 (282) & \asca/SIS &\\
Mar 1995 & $0.5292 \pm 0.03627$ & $5.8\T{-4}$ & BB+CompST & 
	0.974 (279) & 1.029 (281) & \asca/SIS &\\
Apr 1997 & $0.8710 \pm 0.1429$ & $3.0\T{-4} $ & BB+PL & 
	1.321 (78) & 1.335 (80) & \sax/LECS &\\
Apr 1997 & $0.8744 \pm 0.1281$ & $3.1\T{-4} $ & BB+CompST & 
	1.341 (77) & 1.357 (79) & \sax/LECS & \\
\enddata
\tablenotetext{a}{``CompST'' abbreviates the Sunyaev-Titarchuk
Comptonization model, as implemented in XSPEC.}
\end{deluxetable}

\clearpage

\end{document}